# Developing a Magnetism Conceptual Survey and Assessing Gender Differences in Student Understanding of Magnetism

Jing Li and Chandralekha Singh

*Department of Physics and Astronomy, University of Pittsburgh, Pittsburgh, PA 15260, USA*

**Abstract.** We discuss the development of a research-based conceptual multiple-choice survey related to magnetism. We also discuss the use of the survey to investigate gender differences in students' difficulties with concepts related to magnetism. We find that while there was no gender difference on the pre-test, female students performed significantly worse than male students when the survey was given as a post-test in traditionally taught calculus-based introductory physics courses (similar results in both the regular and honors versions of the course). In the algebra-based courses, the performance of the female students and the male students has no statistical difference in the pre-test or the post-test.

**Keywords:** magnetism, gender difference, validity, KR-20, difficulty index, point-biserial coefficient.

**PACS:** 01.40.Fk

## INTRODUCTION

Research-based multiple-choice tests can be useful tools for surveying student learning in physics courses. They are easy and economical to administer and to grade, have objective scoring, and are amenable to statistical analysis that can be used to compare student populations or instructional methods. A major drawback is that the thought processes are not revealed by the answers alone. However, when combined with student interviews, well-designed tests are powerful tools for educational assessment. A number of multiple-choice tests have been developed and widely used by physics instructors to measure students' conceptual learning in physics courses. A commonly used research-based multiple-choice test for mechanics is the Force Concept Inventory (FCI) [1]. In Electricity and Magnetism (E&M), the CSEM and BEMA surveys have been developed which cover E&M concepts discussed in introductory courses [2-3].

Magnetism is an important topic in introductory physics. We developed a research-based 30 item multiple-choice test on magnetism (called the Magnetism Conceptual Survey or MCS) to explore the difficulties students have in interpreting magnetism concepts and in correctly identifying and applying them in different situations. We also wish to know the extent to which the difficulties are universal, and if there is a correlation with instructor or student preparation and background, e.g., whether they are in the calculus-based or algebra-based courses or whether they are females or males. The identification of student difficulties with magnetism for these various groups can help in designing instructional tools to address the difficulties. In this paper, we will focus on gender differences in students' difficulties with magnetism after we summarize the development of the survey including issues related to its validity and reliability.

Previous research shows that there is often a gender difference in student performance in mathematics and other disciplines [4-5] as well as in physics [6-10] which can sometimes be reduced by carefully designed curricula. Here, we explore gender difference in student understanding of magnetism concepts covered in introductory physics courses by surveying students in the calculus- and algebra-based courses using the MCS as a pre-test and a post-test (before and after instruction in relevant concepts).

## MCS SURVEY DESIGN

The Magnetism Conceptual Survey (MCS) covers topics in magnetism discussed in a traditional calculus- or algebra-based introductory physics curriculum up to Faraday's law. During the test design, we paid particular attention to the important issues of reliability and validity [3]. Reliability refers to the relative degree of consistency in scores between testing if the test procedures are repeated in immediate succession for an individual or group. On a reliable survey, students with different levels of knowledge of the topic covered should perform according to their mastery. In this paper, we use the data collected to perform statistical tests to ensure that the survey is reliable within the classical test theory. For example, the reliability index measures the internal consistency of the whole test [3]. One commonly used index of reliability is KR-20 which is calculated for the survey as a whole [3].

Validity refers to the appropriateness of the test score interpretation [3]. A test must be reliable for it to be valid for particular use. The design of the MCS test began with the development of a test blueprint that provided a framework for planning decisions about the desired survey attributes. We tabulated the scope and extent of the content covered and the level of cognitive complexity desired. During this process, we consulted with several faculty members who teach introductory E&M courses routinely about concepts they believed their students should know about magnetism.

We classified the cognitive complexity using a simplified version of Bloom's taxonomy: specification of knowledge, interpretation of knowledge and drawing inferences, and applying knowledge to different situations. Then, we outlined a description of conditions/contexts within which the various concepts would be tested and a criterion for good performance in each case. The tables of content and cognitive complexity along with the criteria for good performance were shown to three physics faculty members at the University of Pittsburgh (Pitt) for review. Modifications were made to the weights assigned to various concepts and to the performance criteria based upon the feedback from the faculty about their appropriateness. The performance criteria were used to convert the description of conditions/contexts within which the concepts would be tested to make free-response questions. These questions required students to provide their reasoning with the responses.

The multiple-choice questions were then designed. The responses to the free-response questions and accompanying student reasoning along with individual interviews with a subset of students guided us in the design of good distracter choices for the multiple-choice questions. In particular, we used the most frequent incorrect responses in the free-response questions and interviews as a guide for making the alternative distracter choices. Four alternative choices have typically been found to be optimal, and we chose the four distracters to conform to the common difficulties to increase the discriminating properties of the items. Three physics faculty members were asked to review the multiple-choice questions and comment on their appropriateness and relevance for introductory physics courses and to detect ambiguity in item wording. They went over several versions of the survey to ensure that the wording was not ambiguous. Moreover, several introductory students were asked to answer the survey questions individually in interviews to ensure that the questions were not misinterpreted.

## MCS ADMINISTRATION

The final version of the MCS was administered both as a pre-test and a post-test to a large number of students at Pitt. These students were from three traditionally taught algebra-based classes, and eight regular (in contrast to the honors) calculus-based introductory classes. In our analysis presented here for the reliability index KR-20, the item difficulty and discrimination indices, and point biserial coefficient of the items, we kept only those students who took the survey both as a pre-test and a post-test except in one algebra-based class. In that class, most students who worked on the survey did not provide their names and seven more students participated in the post-test than the pre-test. Thus, in the algebra-based course, 267 students took the pre-test, and 273 students took the post-test. In the regular calculus-based courses, 575 students took both the pre-test and the post-test.

Pre-tests were administered in the first lecture or recitation at the beginning of the semester in which students took introductory second semester physics with E&M as a major component. The students were not allowed to keep the survey. Post-tests were administered in the recitations after instruction in all relevant concepts on magnetism covered in the MCS. Students were typically asked to work on the survey for a full class period (40-50 minutes).

The KR-20 for the combined algebra-based and calculus-based data is 0.83, which is reasonably good by the standards of test design [3]. The MCS was also administered to 42 physics graduate students enrolled in a first year course for teaching assistants to bench mark the performance that can be expected of the undergraduate students. The average score for the graduate students is 83% with a KR-20 of 0.87.

The item difficulty is a measure of the difficulty of a single test question [3]. It is calculated by taking the ratio of the number correct responses on the question to the total number of students who attempted to answer the question. Figure 1 shows the difficulty index for each item in the survey for the sample of 848 students obtained by combining the algebra-based and calculus-based classes. The average difficulty index is 0.46 which falls within the desired criterion range [3]. The average difficulty index for the algebra-based class is 0.45 which is lower than 0.53 for the calculus-based class.

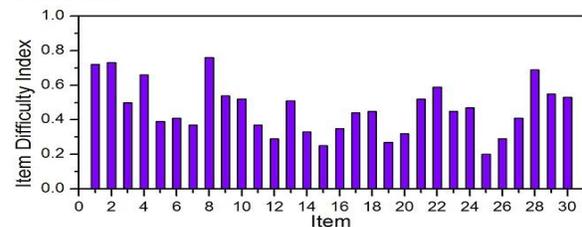

**FIGURE 1.** Difficulty index for various items in the MCS

The item discrimination index measures the discriminatory power of each item in a test [3]. A majority of the items in a test should have relatively high discrimination indices to ensure that the test is capable of distinguishing between strong and weak mastery of the material. A large discrimination index for an item indicates that students who performed well on the test overall performed well on that item. The average item discrimination index for the combined 848 students sample including all items on the MCS is 0.33 which is reasonable from the standards of test design [3]. Figure 2 shows that for this sample the item discrimination indices for 22 items are above 0.3. The average discrimination index for the algebra-based class is 0.29 and for the calculus-based class it is 0.33.

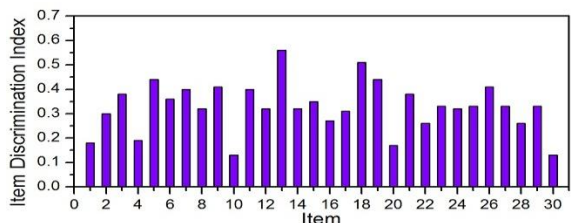

**FIGURE 2.** Discrimination index for the MCS items

The point biserial coefficient is a measure of consistency of a single test item with the whole test [3]. It is a form of a correlation coefficient which reflects the correlation between students' scores on an individual item and their scores on the entire test. The widely adopted criterion for a reasonable point biserial index is 0.2 or above [3]. The average point biserial index for the MCS is 0.42. Figure 3 shows that all items have a point biserial index equal to or above 0.2.

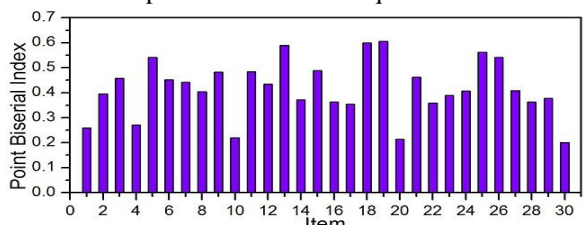

**FIGURE 3.** Point biserial coefficient for the MCS items

## PERFORMANCE BY GENDER

For analyzing gender difference in students' performance on the MCS, we separate our data into male and female groups. Only the students who provided this information were kept in this analysis. The gender comparison in the algebra-based classes includes 121 females and 110 males (total 231 students) on the pre-test and 106 females and 91 males (total 197) on the post-test. There were 168 females and 403 males (total 571 students) from the regular (not honors) calculus-based classes who took both the pre-test and the post-test and are included in the analysis below. In addition to comparing the results from the algebra-based and regular calculus-based classes, we also analyzed the gender data for the post-test of 95 students enrolled in the honors calculus-based introductory physics course. The honors students were not administered the MCS as a pre-test.

We perform analysis of variance (ANOVA) to investigate the gender differences from the pre-test and the post-test MCS data. Our null hypothesis is that there is no significant gender difference on MCS. If the p-value is less than the significance level 0.05, the rule of thumb is to conclude that the assumption is false (here it will imply that there is a significant difference between the male and female performance).

Tables 1-2 show the results for the algebra-based students on the pre-test and the post-test. Table 1 shows that in the pre-test, the means are 7.3 and 7.1 for the males and the females respectively. The p-value, 0.942, which is larger than 0.05, suggests no significant difference between the males and females on the pre-test in algebra-based classes. Table 2 shows the results for the post-test. It shows that the mean for the females is 12.3 compared to the mean for the males, 13.2. The p-value, 0.355, suggests that even on the post-test, the algebra-based students do not have a significant difference in performance based on gender.

**TABLE 1.** Algebra-based course pre-test performance by gender

| Gender | N | Mean | S.D. | P value |
|---|---|---|---|---|
| Male | 91 | 7.3 | 2.50 | 0.942 |
| Female | 106 | 7.1 | 2.36 | |

**TABLE 2.** Algebra-based course post-test performance by gender

| Gender | N | Mean | S.D. | P value |
|---|---|---|---|---|
| Male | 110 | 13.2 | 5.20 | 0.355 |
| Female | 121 | 12.3 | 5.57 | |

The results for the regular calculus-based classes are qualitatively different from the algebra-based classes for the post-test. The pre-test mean for males is 8.5 and for females is 7.8. The p-value for analysis of variance between these groups is 0.490 suggesting no significant difference based on gender on the pre-test. However, the results shown in Table 4 suggest that there is a significant difference on the post-test and males outperformed females. The mean for the males is 15.3 compared to the mean for the females which is 13.0 ( p-value, 0.019).

**TABLE 3.** Regular calculus-based course pre-test performance by gender

| Gender | N | Mean | S.D. | P value |
|---|---|---|---|---|
| Male | 403 | 8.5 | 3.40 | 0.490 |
| Female | 168 | 7.8 | 3.05 | |

**TABLE 4.** Regular calculus-based course post-test performance by gender

| Gender | N | Mean | S.D. | P value |
|---|---|---|---|---|
| Male | 403 | 15.3 | 6.20 | 0.019 |
| Female | 168 | 13.0 | 5.38 | |

**TABLE 5.** Honors calculus-based course post-test performance by gender

| Gender | N | Mean | S.D. | P value |
|---|---|---|---|---|
| Male | 75 | 17.4 | 5.89 | 0.030 |
| Female | 20 | 14.1 | 6.21 | |

The gender difference also exists on the post-test for the calculus-based honors introductory physics course. Table 5 shows that the mean for 75 males is 17.4 and for 20 females is 14.1 (p-value is 0.030).

To summarize the data presented in Tables 1-5, for both the algebra- and calculus-based classes, there is no significant difference between the males and females on the pre-test. After traditional instruction, there is still no gender difference in the algebra-based classes. However, a statistically significant difference appeared on the post-test for the calculus-based classes in which there are significantly fewer females in each class than males (both regular and honors).

**TABLE 6.** Percentage of correct response on each item by gender in algebra- and regular calculus-based courses

| Item | Alg-M | Alg-F | Calc-M | Calc-F |
|---|---|---|---|---|
| 1 | 61 | 55 | 79 | 73 |
| 2 | 62 | 55 | 82 | 74 |
| 3 | 40 | 39 | 56 | 53 |
| 4 | 59 | 50 | 73 | 67 |
| 5 | 36 | 43 | 42 | 38 |
| 6 | 49 | 50 | 38 | 39 |
| 7 | 46 | 42 | 38 | 28 |
| 8 | 79 | 74 | 80 | 71 |
| 9 | 44 | 40 | 62 | 56 |
| 10 | 48 | 41 | 57 | 50 |
| 11 | 36 | 32 | 41 | 37 |
| 12 | 28 | 20 | 35 | 26 |
| 13 | 37 | 32 | 61 | 53 |
| 14 | 25 | 18 | 42 | 24 |
| 15 | 25 | 26 | 28 | 23 |
| 16 | 21 | 8 | 45 | 43 |
| 17 | 41 | 54 | 42 | 48 |
| 18 | 44 | 30 | 56 | 38 |
| 19 | 33 | 25 | 30 | 21 |
| 20 | 25 | 31 | 36 | 29 |
| 21 | 51 | 55 | 54 | 51 |
| 22 | 64 | 52 | 63 | 51 |
| 23 | 39 | 32 | 54 | 39 |
| 24 | 55 | 46 | 50 | 38 |
| 25 | 18 | 20 | 24 | 14 |
| 26 | 28 | 32 | 33 | 23 |
| 27 | 37 | 35 | 47 | 36 |
| 28 | 72 | 69 | 72 | 63 |
| 29 | 66 | 67 | 54 | 46 |
| 30 | 49 | 55 | 54 | 52 |

We looked at students' responses to each MCS item individually to understand how males and females performed on each question. The results are shown in Table 6. Table 6 shows that in the algebra-based classes, males outperformed females on 20 questions including 4 questions on which the differences are larger than 10%. On the 10 questions on which females outperformed males, only one has a difference of more than 10%. In the calculus-based classes, males outperformed females on 28 questions and 9 of them have a difference larger than 10%. On the other two questions, females only performed slightly better than the males.

Answering many of the questions on the MCS correctly requires that students be able to visualize the situation in three dimensions (3D). For example, some questions require that students apply the right hand rule to figure out the directions of the magnetic field or the force on a moving charge or a current carrying wire. Some prior research suggests that females generally have a better verbal ability but worse spatial ability than males which can restrict their reasoning in 3D and often there is a correlation between students' spatial ability and their self-confidence [11-12]. The reasons for gender differences are quite complex and can include factors such as accumulated societal bias.

## SUMMARY

We developed and administered the MCS as a pre-test and a post-test in the introductory physics classes. We find no gender difference in the algebra-based courses but a significant gender difference in both the regular and honors calculus-based courses on the post-tests (but not on the pre-tests). Further research is needed to investigate the reasons for these differences.

## REFERENCES

1. D. Hestenes, M. Wells, and G. Swackhamer, Phys. Teach. **30**, 141 (1992).
2. D. Maloney, T. O'Kuma, C. Hieggelke, and A. Van Heuvelen, Am. J. Phys. **69**, S12 (2001).
3. L. Ding, R. Chabay, B. Sherwood, and R. J. Beichner, Phys. Rev. ST PER **2(1)**, 7 (2006).
4. J. S. Hyde, E. Fennema and S. J. Lamon, Psy. Bulletin, **107(2),** 139-155 (1990).
5. J. Kahle, and J. Meece in *Handbook of research on science teaching*, edited by D. Gabel, New York: McMillan, 542, 1994.
6. M. Lorenzo, C. Crouch and E. Mazur, Am. J. Phys. **74(2),** 118, (2006)**.**
7. S. Pollock, N. Finkelstein and L. Kost, Phys. Rev. ST PER **3**, 010107 (2007).
8. L. Kost et al., Phys. Rev. ST PER **5**, 010101 (2009).
9. P. Kohl, H. Kuo, PERC Proceedings. **1179**, 173 (2009).
10. L. Kost et al., PERC Proceedings, **1179**, 177 (2009).
11. D. Law, J. Pellegrino and E. Hunt, Psychological Science, **4(1),** 35 (1993).
12. M. Beth Casey, R. Nuttall and E. Pezaris, Journal for Research in Mathematics Education, **32(1),** 28 (2001).